# AutoLogger: A Multi-Agent Framework for the End-to-End Automated Logging

Renyi Zhong
The Chinese University of Hong Kong
Hong Kong, Hong Kong
ryzhong22@cse.cuhk.edu.hk

Yintong Huo
Singapore Management University
Singapore, Singapore
ythuo@smu.edu.sg

Wenwei Gu
Nankai University
Tianjin, China
wwgu@nankai.edu.cn

Yichen Li*
The Chinese University of Hong Kong
Hong Kong, Hong Kong
ycli21@cse.cuhk.edu.hk

Michael R. Lyu
The Chinese University of Hong Kong
Hong Kong, Hong Kong
lyu@cse.cuhk.edu.hk

## Abstract

Software logging is critical for system observability, yet developers face a dual crisis of costly overlogging and risky underlogging. Existing automated logging tools often overlook the fundamental whether-to-log decision and struggle with the composite nature of logging. In this paper, we propose AutoLogger, a novel hybrid framework that addresses the complete the end-to-end logging pipeline. AutoLogger first employs a fine-tuned classifier, the Judger, to accurately determine if a method requires new logging statements. If logging is needed, a multi-agent system is activated. The system includes specialized agents: a Locator dedicated to determining where to log, and a Generator focused on what to log. These agents work together, utilizing our designed program analysis and retrieval tools. We evaluate AutoLogger on a large corpus from three mature open-source projects against state-of-the-art baselines. Our results show that AutoLogger achieves 96.63% F1-score on the crucial whether-to-log decision. In an end-to-end setting, AutoLogger improves the overall quality of generated logging statements by 16.13% over the strongest baseline, as measured by an LLM-as-a-judge score. We also demonstrate that our framework is generalizable, consistently boosting the performance of various backbone LLMs.

*Corresponding author.

## 1 Introduction

Software logging is the cornerstone of observability in distributed systems, enabling failure diagnosis, performance optimization, and security auditing [4, 8]. However, modern software engineering faces a dual logging crisis. Log volumes are growing at 250% year-over-year [28]. Organizations spend seven to eight figures annually on logging infrastructure while suffering up to 3× performance overhead [44]. Significant cost savings are realized by simply eliminating unnecessary logs [32]. Conversely, insufficient logging creates severe vulnerabilities. The Uptime Intelligence Institute reports 78% of outages as preventable with better monitoring and observability [13, 26]. This dual crisis of simultaneous overlogging and underlogging demands a fundamental rethinking of logging decisions.

Effective logging is a deceptively complex task, proving to be a persistent pain point even for experienced developers [9]. It demands sophisticated predictive reasoning, as developers must anticipate future diagnostic needs long after deployment [43]. This decision is multi-dimensional, spanning whether-to-log *(is logging necessary?)*, where-to-log *(which code location?)*, and what-to-log *(which level, variables, and message?)* [5, 17]. Empirical studies confirm this human challenge, revealing high rates of inconsistent severity levels (35.4%) [44, 46] and frequent log deletions (32.1%) [11], which suggest low developer confidence. This inherent complexity has motivated substantial research on automated assistance. Early approaches employed traditional machine learning with extracted features [48] or information retrieval techniques [45] to address isolated sub-problems like location prediction [16] or message generation [5]. The deep learning era brought end-to-end solutions like LANCE [23], LEONID [22], and FastLog [39]. The emergence of LLMs has further advanced the field with tools like UniLog [40] and SCLogger [15].

Despite continuous improvements, existing automated logging approaches may suffer from two fundamental limitations that constrain their practical effectiveness. First, they cannot be applied as true end-to-end solutions because they rely on the unrealistic assumption that the crucial whether-to-log decision has already been made. As our analysis of manual logging challenges confirms, developers face a difficult trade-off between capturing sufficient information and overwhelming system resources [42, 48]. This whether-to-log decision is a primary cognitive hurdle, yet it is almost universally ignored by modern tools. Nearly all existing state-of-the-art tools are architected as code generation systems. This design oversight means they are incapable of identifying methods that developers may have overlooked (i.e., underlogging). Even more concerning, when these are used without discretion, they intensify the overlogging aspect of the dual crisis. This happens because they do not have a system to check whether a method is already logging-saturated or if logging isn't necessary at all.

Second, these approaches are architecturally ill-suited for the fact-intensive reasoning required by logging. Effective logging is not a monolithic generation task. It is a composite of distinct subtasks demanding different reasoning capabilities. A model must: (1) precisely locate semantically-critical execution points, requiring deep control-flow understanding [16]; (2) accurately select diagnostic variables, necessitating precise data-flow tracing [20]; and (3) correctly generate content, demanding adherence to project-specific styles [5].

Monolithic models, including modern LLMs, attempt to solve all these sub-tasks in a single, static forward pass. This creates a gap between the model's static, pre-trained knowledge and the factual grounding required by the specific code context. This mismatch forces the model to guess when faced with uncertainty, leading directly to hallucination. Lacking a tool to actively verify in-scope variables, an LLM may hallucinate a variable name or logging one that is out of scope. Lacking a mechanism to analyze precise control-flow, it may place a log in a syntactically plausible but semantically useless location. Lacking the ability to retrieve project conventions, it may generate text that violates established logging styles. This architectural flaw produces these sub-optimal results. It forces one static model to be an expert in program analysis, data-flow tracing, and stylistic generation without interactive, fact-checking capabilities.

This limitation motivates a shift toward a more dynamic architecture. A multi-agent system (MAS) [7] provides such an alternative. A MAS decomposes the composite problem. It assigns distinct subtasks to specialized agents. These agents are not limited to their pre-trained knowledge. They can be empowered with a pool of tools, enabling interactive code probing. This model grants agents autonomy. They can decide whether tool-use is necessary. For simple methods, agents can provide a result directly. Complex, tool-based analysis is reserved for ambiguous or fact-intensive cases. This dynamic, tool-assisted reasoning directly counters the limitations of monolithic models. Agents can analyze control flow, verify variable scope, and retrieve stylistic conventions. This process grounds the generation in verifiable facts and mitigates hallucination.

To systematically address these limitations, we propose AutoLogger, a novel hybrid framework built on task decoupling and component specialization. AutoLogger operates in two stages. First, a fine-tuned classifier, the Judger, confronts the neglected whether-to-log decision, acting as an efficient filter to identify logging-unsaturated methods. If logging is deemed necessary, the framework activates its second stage: a MAS that decomposes the composite generation task. This MAS features specialized agents, a Locator for the where-to-log task and a Generator for the what-to-log task. These agents collaborate using a pool of tools of program analysis (e.g., backward slicing, variable extraction) and retrieval utilities. This design grounds the agents in factual code analysis, mitigating LLM hallucination and ensuring stylistic consistency with the existing codebase.

We evaluate AutoLogger on a large corpus from three popular mature open-source projects. Our results show that the specialized Judger achieves a 96.63% F1-score on the crucial whether-to-log decision, far surpassing general-purpose LLM baselines. In a comprehensive end-to-end evaluation, AutoLogger improves the overall quality of generated logging statements by 16.13% over the strongest baseline, as measured by an LLM-as-a-judge score. We also demonstrate that our framework is generalizable, consistently boosting the performance of various backbone LLMs, which confirms the value of our hybrid, agent-based architecture.

In summary, this paper makes the following contributions.

- To the best of our knowledge, we are the first to systematically address the complete automated logging pipeline by formally introducing and solving the neglected whether-to-log task.
- We propose AutoLogger, a novel hybrid framework that combines a fine-tuned classifier (the Judger) with a MAS. This MAS decomposes the composite generation task, using specialized Locator and Generator agents that collaborate with a pool of tools of program analysis utilities to ground their reasoning.
- We conduct a comprehensive evaluation on a large-scale corpus from three mature popular open-source project. Experimental results show that AutoLogger outperforms existing approaches and is adaptable to different backbones.
- All the code and data used in this paper is publicly available in our replication package [1].

## 2 Methodology

### 2.1 Overview

We first define the problem of automated logging. Given a method's source code as input (i.e., the target method), this task aims to address the complete logging pipeline. This pipeline involves three core sub-tasks: (1) whether-to-log: determining if the method requires any logging; (2) where-to-log: predicting the specific code line number for log insertion if logging is needed; and (3) what-to-log: generating a complete logging statement, including its level, text, and variables, for the specified location. The goal of AutoLogger is to address all three sub-tasks systematically.

We propose AutoLogger, a hybrid framework combining a fine-tuned model with an LLM-based multi-agent system, as illustrated in Figure 1. The existing approach with individual LLM struggles to balance the distinct objectives of classification and generation. AutoLogger solves this by decoupling the process into two stages. In the first stage, the target method is processed by the **Judger** component. The Judger is a fine-tuned binary classification model specialized to efficiently determine if the method requires logging, outputting a *LOG* or *NO_LOG* decision. If the Judger decides to log, the framework activates the second stage: a MAS. This system consists of two collaborating, specialized agents to handle the generation task. The **Locator** addresses the where-to-log problem by analyzing the method's structure to identify suitable line for logging. For location identified by the Locator, the **Generator** is invoked to address the what-to-log problem. To ensure high-quality generation, the agents are supported by a pool of tools containing program analysis utilities. This toolset provides essential context, such as relevant variables, data-flow dependencies, and existing logging styles from the project. Both agents leverage this comprehensive context to create the final, complete logging statement for the specified location.

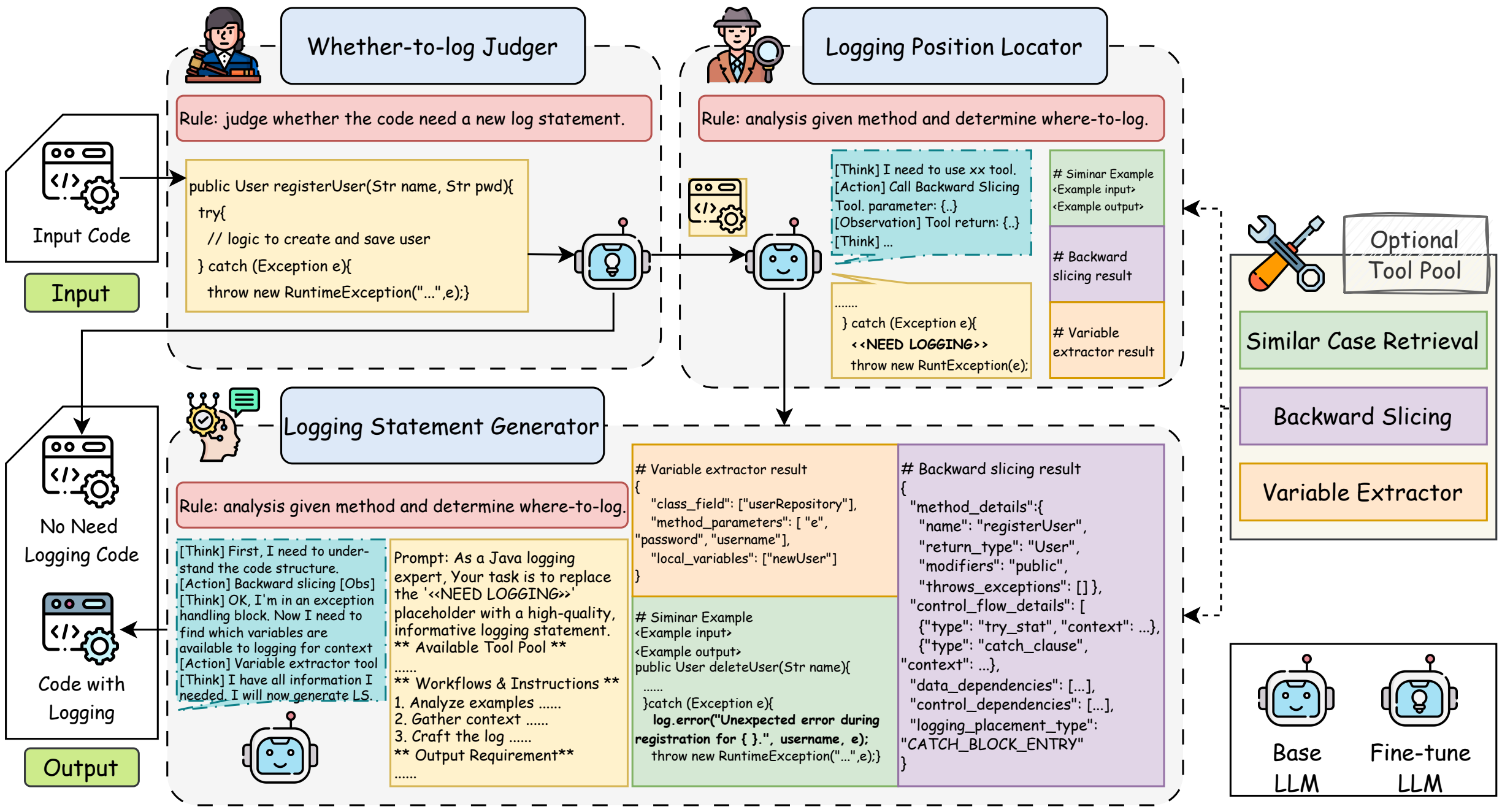

**Figure 1: The framework of AutoLogger.**

## 2.2 Stage I: Determine Logging Necessity via Judger

The Judger component is the first stage of the AutoLogger. Its primary objective is to address the "whether-to-log" decision. This task moves beyond a simple check for existing logs. The Judger is specialized to identify methods that are logging-unsaturated, meaning they still require new logging statements. This component acts as an efficient, preliminary filter. It prevents the system from wasting computational resources on methods that are already logging-saturated or do not need logs. Only methods that require logging proceed to the resource-intensive multi-agent generation stage.

We deliberately chose a fine-tuned classifier for this role instead of an LLM-based agent for two key reasons. First, the whether-to-log decision is a specialized classification task that is highly dependent on repository-specific conventions. Effective logging is not a universal constant; it is guided by implicit, project-level standards, practices, and logging guides. A general-purpose LLM agent may apply generic logging knowledge that conflicts with these local norms (e.g., a project's specific rules about logging in getters/setters or inside loops). In contrast, a fine-tuned model is optimized to learn and internalize these subtle, repository-specific patterns directly from the project's data, allowing it to act as a highly accurate expert on this single, repetitive decision. Second, this component must be highly efficient. As a preliminary filter, the Judger processes all methods. A fine-tuned model offers a fast, low-latency, and cost-effective solution, whereas a full LLM agent-based approach may introduce computational overhead, making it impractical as a gatekeeper for the more resource-intensive MAS.

We formulate this problem as a binary classification task. The model takes the full source code of a Java method as input and outputs a single *LOG* or *NO_LOG* label. A *LOG* label indicates the method needs a new log, while *NO_LOG* indicates the method is logging-saturated or requires no logging. The training data for the Judger is constructed using a specific strategy. Positive samples represent methods needing logs. We create them by taking a method with $n(n \geq 1)$ logs and removing one, leaving n-1 logs. This process simulates a method where a log is missing and logging-unsaturated. Negative samples are the original, unmodified methods from our corpus. These negative samples represent a stable state, either because they require no logs or because their existing logs are already sufficient. Therefore, the Judger learns to distinguish between logging-saturated methods and logging-unsaturated methods.

We selected Qwen2.5coder-14b as the backbone model for the Judger due to its strong code comprehension capabilities. However, fully fine-tuning a 14-billion parameter model is computationally prohibitive. To balance performance and parameter efficiency, we employ a Parameter-Efficient Fine-Tuning (PEFT) strategy, specifically Low-Rank Adaptation (LoRA) [10]. LoRA drastically reduces the number of trainable parameters, lowers memory requirements, and produces a task-specific model without catastrophic forgetting.

To further enhance the Judger's training, we formulate the classification task as an instruction-following problem. We do not feed the raw code directly to the model. Instead, we construct a context-rich prompt for each training sample. We use a BM25 retrieval algorithm [31] to search the valid corpus and find a method that is syntactically similar to the target method. This similar case and its corresponding ground-truth label are then used as a one-shot example within the prompt. The final prompt fed to the model includes this in-context example, the target method's code, and a clear instruction to classify the target. This strategy allows the model to learn by comparing the target method to a relevant example, improving its ability to discern complex logging patterns.

## 2.3 Stage II: Multi-Agent Systems

*2.3.1 MAS Overview and Rationale.* This second stage of AutoLogger activates for methods that the Judger classifies as *LOG*. The primary objective of this stage is to comprehensively solve the where-to-log and what-to-log problems. We determined that a single, monolithic model is ill-suited for this process. Generating a high-quality logging statement is not one task but a composite of distinct sub-tasks, including control-flow analysis, data-flow tracing, style retrieval, and text generation. A single model struggles to optimize these competing concerns simultaneously. Therefore, we designed a MAS based on the principles of task decomposition and agent specialization. This MAS consists of three core components. First, we use Specialized Agents, which are LLMs tasked with specific roles: the Locator finds logging location, and the Generator creates the content of logging statement. Second, a Tool Pool provides these agents with program analysis and retrieval capabilities. Third, an MAS Orchestrator acts as the middleware, managing the workflow and coordinating all communication between the agents and the tools.

*2.3.2 MAS Orchestrator and Collaborative Workflow.* The MAS Orchestrator component is the central middleware of AutoLogger. It orchestrates the communication between the specialized agents with tools and manages the overall workflow. The Orchestrator itself is not an LLM but a stateful procedural component that executes a defined two-phase pipeline. It initiates agent calls, parses their responses, refines tool outputs, and manages the flow of context between agents.

A primary function of the Orchestrator is parsing and refining agent output, which is critical because LLM agent responses can be unstructured. The Orchestrator receives the raw text response from the Locator and applies two key parsing heuristics. First, it uses regular expressions to locate and extract Java code blocks, which are the primary payload. Second, the Orchestrator parses the raw output of tools used by the Locator. For instance, it processes the string output from the similar case retrieval tool. It then reformats this raw data into a structured, LLM-friendly case format. This refinement step translates raw tool data into actionable in-context examples for the following agent.

The Orchestrator also manages the prompt update and state transfer between the two agents. After the Locator executes, the Orchestrator extracts all relevant tool outputs, such as the newly formatted retrieval context. It then dynamically constructs a new, comprehensive prompt for the Generator. This prompt includes the marked-up code from the Locator and the formatted tool outputs. This procedure guarantees that the Generator receives all essential information from the previous stage, reducing the need to repeatedly use those tools and thus saving costs.

Finally, the Orchestrator is responsible for ensuring system robustness and deterministic execution. The workflow is governed by several protective mechanisms. First, a global temporal constraint is enforced on the entire process. This strategy mitigates the risk of non-terminating agent states and ensures predictable termination. Second, the system integrates a comprehensive exception handling strategy. This strategy manages unforeseen runtime failures by attempting to retry the operation at the beginning to ensure resilience against transient errors [2]. Persistent failures that are not resolved by retries will lead to a controlled termination of the process. These mechanisms work together to ensure the reliability of the complex agent coordination process.

*2.3.3 Agent Design.* This MAS is driven by two specialized agents, the Locator and Generator. These agents are the primary executors within the workflow, powered by a default deepseek-v3.1 backbone. We designed them to operate using an autonomous, iterative reasoning model. This interaction model is inspired by the ReAct (Reason-Act) framework [41]. In this paradigm, the agent's response is structured to include both *thoughts* and a *command*. The *thoughts* field contains the agent's internal thinking, which explains its reasoning and current strategy. The *command* field specifies the name of a tool from the tool pool we provided to execute and its corresponding arguments.

This ReAct cycle allows the agents to perform complex tasks. The Orchestrator first parses the agent's *command*, executes the specified tool, and then feeds the tool's output back to the agent. This output is appended to the agent's prompt for the next reasoning cycle. This iterative loop continues until the agent has gathered sufficient information and determines that its sub-task is complete, at which point it produces a final answer instead of a tool command.

The Locator is responsible for the where-to-log problem. It receives the full method code from the Orchestrator. Its prompt instructs it to identify semantically critical execution points for logging. To achieve this, the agent iteratively reasons about the code's structure. It can autonomously decide to call tools that help it analyze the method's control flow and identify key logical blocks. After one or more reasoning cycles, it concludes its task by outputting the method code with the added *«need_logging»* tag to show the result of decided logging location, which the Orchestrator then parses.

The Generator is responsible for the what-to-log problem. The Orchestrator invokes this agent, providing it with the marked-up source code from the Locator. The Generator's task is to find the «need_logging» placeholder and replace it with a complete logging statement. To do this, it initiates a series of ReAct cycles to gather context specific to that placeholder's location. The agent autonomously decides which tools to call in sequence. It might first choose to identify all available variables within that scope. Based on that output, it may then call a different tool to determine which of those variables are most relevant to the program's state. Finally, it may decide to retrieve stylistically similar logging examples. After gathering all necessary context, it composes the final logging statement and outputs the complete source code with the placeholder replaced.

*2.3.4 Tool Design.* The AutoLogger agents' autonomous decisions are grounded in a pool of tools. The design philosophy for these tools is to provide concrete, verifiable information from the source code. This approach grounds the LLM's reasoning in program analysis in order to prevent factual hallucinations and ensure the generated logging statements are contextually accurate and stylistically consistent. We categorize the tools by their primary function.

The first goal is similar context and style retrieval. To ensure agents construct a logging statement that matches the project's existing conventions, we provide our *similar_case_retrieval* tool. This tool enables in-context learning for logging style. It operates on a pre-indexed corpus of (code_before, code_after_logging) example pairs. When invoked with a target code snippet, the tool tokenizes the snippet and uses the BM25 algorithm [31] to retrieve the single most similar example pair from the corpus. This pair is then passed to the agent, guiding it to produce a logging statement that is stylistically consistent with project norms.

The second goal is to leverage program analysis to prevent hallucinations in LLMs. These tools provide a semantic understanding of the code. The *variable_extractor* tool performs static scope analysis. Its motivation is to provide the Generator with a complete set of all available variables, preventing the agent from hallucinating variable names. The tool uses a tree-sitter parser to build an Abstract Syntax Tree (AST). It then executes a set of pre-defined structural queries against the AST to find and return a structured list of all accessible class fields, method parameters, and local variables. To provide a comprehensive analysis of the logging location, we also design a *backward_slicing* tool. This tool uses a target line number as its analysis target. It performs a multi-faceted analysis to build a complete contextual picture. First, it identifies high-level method details, such as its return type, modifiers, and thrown exceptions. It also analyzes the target position to determine the placement type, such as `METHOD_ENTRY` or `CATCH_BLOCK_ENTRY`. Concurrently, it performs a static backward slice from the criterion. This slice identifies two critical sets of information: the data-dependency statements (e.g.,variable assignments that influence the program state), and the control-flow context (e.g., the specific conditions of enclosing if or for statements). This combined analysis provides the agent with all relevant context, from the high-level method signature down to the specific data and control dependencies at the target point.

## 3 Experimental Design

### 3.1 Subject Projects

Our empirical evaluation is grounded in a corpus constructed from three large-scale, mature open-source projects: Doris [35], Kafka [36], and ZooKeeper [37]. These projects were chosen due to their widespread application in industry and their variety of use cases: a high performance database, a distributed messaging platform, and a distributed coordination service. This domain diversity ensures that our evaluation is not biased toward a specific type of software architecture. Table 1 presents detailed statistics for each project, which collectively span from 160.4k to 1.09M lines of code (LOC). Each system contains between 1.9k and 3.4k logging statements, and all possess a development history exceeding a decade, providing a rich and evolved codebase for analysis. It's important to mention that we chose not to use other common logging datasets [15, 23, 33] because such datasets usually include only positive instances, meaning they contain code segments already marked as needing logging. This method assumes logging is required, which doesn't allow us to assess a model's capability in making the core decision of whether-to-log.

**Table 1: Details of studied projects.**

| Project | Version | #Stars | #Forks | #LOC | #LOLS |
|---|---|---|---|---|---|
| Doris | 3.0.5 | 13.6k | 4.8k | 845.3k | 2.9k |
| Kafka | 3.9.1 | 30.0k | 14.3k | 1.09M | 3.4k |
| ZooKeeper | 3.9.4 | 12.5k | 7.3k | 160.4k | 1.9k |

To ensure the integrity of our experimental results, we implemented rigorous measures to mitigate the potential data leakage. First, to prevent information from leaking between our dataset splits, we enforced a separation at the file level. All methods extracted from the same Java file were exclusively assigned to the same dataset partition [47]. To counter the possibility of pretraining contamination, which could occur if the LLM had been exposed to the source code of these well-known projects during its original pre-training, we implemented two specific transformations. We first reformatted the entire codebase using the Google Java Formatter [6] to standardize and alter the original code style. Subsequently, we encapsulated each method within a generic wrapper class (Class A), effectively stripping the original context at the class level that the model might recognize [33].

We employed a mixed-project splitting strategy. The final corpus, comprising data from all three subject projects, was partitioned into training, validation, and testing sets. We employed stratified sampling [25] to ensure that the distribution of methods with and without logging statements was consistent across all three splits, thereby enabling a fair and balanced evaluation. The final training set contains 33,247 instances, the validation set contains 8,312 instances, and the test set contains 2,242 instances.

### 3.2 Baselines

From a review of relevant studies published in SE venues, we select the following baselines: **UniLog** [40], **FastLog** [39], and **SCLogger** [15]. Note that LANCE [23] and LEONID [22] are excluded from our comparison, as their underlying CodeT5-based models impose a 512-token maximum input length. This limitation is incompatible with our dataset, which is designed to preserve real-world code distributions without such input length restrictions. Additionally, we examine several LLM baselines, including general-purpose LLMs (**Claude-sonnet-4**, **GPT5**, **Deepseek-v3.1**), and a thinking LLM (**Deepseek-R1**).

## 3.3 Evaluation Metrics

*3.3.1 Whether-to-log.* We use **Precision** to measure the proportion of correctly identified logging locations among all predicted locations, and **Recall** to measure the proportion of actual logging locations that our model successfully identifies. The **F1-score**, as the harmonic mean of Precision and Recall, provides a unified measure that balances these two aspects. Furthermore, because logging data is inherently imbalanced—with far more code lines without logs than with them—we also report **Balanced Accuracy (BA)**. BA offers a more robust performance measure by averaging the recall of each class, thereby preventing the model's performance on the majority (non-logged) class from skewing the overall evaluation. BA is calculated as follows: $BA = \frac{1}{2} \times \frac{TP}{TP+FN} + \frac{1}{2} \times \frac{TN}{TN+FP}$.

*3.3.2 Where-to-log.* We use **Position Accuracy (PA)** to evaluate the performance of logging location. To quantify PA, we compare the predicted locations of logging statements against their ground truth positions in the source code. Same as related work [15], we count a prediction as correct (PA = 1) if the predicted line number deviates from the true line number by at most one and both lines are located within the same code block.

*3.3.3 What-to-log. a) Logging Level.* We use the **Level Accuracy (LA)** and **Average Ordinal Distance Score (AOD)** for evaluating the prediction of logging levels. LA is defined by the formula where you divide the number of accurately predicted log levels ($LS_{level_correct}$) by the total count of logging statements ($LS_{all}$), represented as $\frac{LS_{level_correct}}{LS_{all}}$. AOD measures how well the current logging level matches the recommended level for each logging statement, as outlined in [17]. The AOD can be calculated using the formula: $AOD = \frac{\sum_{i=1}^{N}(1-\frac{Dis(a_i,s_i)}{MaxDis(a_i)})}{N}$, where $N$ is the number of logging statements being assessed. Here, $MaxDis(a_i)$ indicates the greatest possible distance for the actual log level $a_i$.

*b) Logging Variable.* We evaluate dynamic logging variables using **Precisely Match Rate (PMR)** and the **Variable-F1**. PMR measures the consistency of variable capture, defined as the ratio of logging statements with exactly matched variables ($LS_{variable_correct}$) to the total number of logging statements ($LS_{all}$): $PMR = \frac{LS_{variable_correct}}{LS_{all}}$ The Variable-F1 provides a combined assessment by calculating the harmonic mean of precision and recall. These metrics are determined based on the set of predicted variables ($S_{ud}$) and the set of ground-truth variables ($S_{gt}$): The evaluation metrics are defined as follows: $Precision = \frac{|S_{ud} \cap S_{gt}|}{|S_{ud}|}$, $Recall = \frac{|S_{ud} \cap S_{gt}|}{|S_{gt}|}$, $Variable - F_1 = 2 \cdot \frac{Precision \cdot Recall}{Precision + Recall}$.

*c) Logging Text.* Same as previous study [5, 14, 23] , our evaluation of static logging texts is conducted through the application of two metrics commonly employed in the domain of machine translation: BLEU [27] and ROUGE [18]. They provide a normalized score continuum from 0 to 1, with elevated scores indicative of a closer resemblance. Specifically, we use BLEU-4 and ROUGE-L to compare the overlap of n-grams.

Similar to the related work [15, 47], we compute metrics for what-to-log exclusively when the predictions for whether-to-log and where-to-log are accurate. This approach ensures meaningful evaluation by only analyzing log content when the system correctly identifies both the need for logging and the appropriate location.

*3.3.4 Overall Quality.* To evaluate the overall quality of the generated logging statement, we employ an 'LLM-as-a-judge' methodology [38]. Our evaluation framework utilizes an ensemble of three state-of-the-art, code-proficient LLMs: Deepseek-v3.1, Deepseek-R1, and GPT5. Each LLM judge assesses a generated logging statement by considering its source code context, the generated output, and the corresponding ground truth. Based on a comprehensive scoring rubric [1], the judges assign a holistic **LLMJudge-Score** from 0 to 3, evaluating the accuracy of the logging placement, verbosity level, static text, and dynamic variables. The final score for each statement is the average of the scores from the three LLM judges.

## 3.4 Implementation

To evaluate existing automated logging approaches, we reproduced conventional methods using replication packages provided by their authors. We infer the LLMs by calling their official APIs. To ensure reproducibility, we set the temperature to 0. For the fine-tuning part, we employed the Axolotl framework. All SOLMs were fine-tuned for one epoch. We use a learning rate of 1e-4 with a cosine scheduler, a global batch size of 64, and the Adam optimizer [12]. We use LoRA with a rank (r) of 16 and an alpha of 32 to save memory. All experiments, including fine-tuning and inference, were conducted on a single NVIDIA A100 80GB GPU provided by Modal [24], a serverless cloud infrastructure platform. For our prompt strategy, for the whether-to-log task, all models (both our fine-tuned Judger and the LLM baselines) utilized an identical prompt. This prompt was augmented using BM25 to retrieve the most similar case from the valid corpus as a one-shot in-context learning example. The exact prompt templates are detailed in our replication package [1].

# 4 Experimental Result

## 4.1 Research Questions (RQ)

- **RQ1: (Overall Effectiveness)** To what extent does AutoLogger outperform SOTA methods in automated logging?
- **RQ2: (Contribution of Tools)** How much does each tool we designed contribute to AutoLogger's effectiveness?
- **RQ3: (Generalizability)** How generalizable is AutoLogger for different backbone models?

## 4.2 RQ1: Overall Effectiveness

*4.2.1 Motivation.* The primary objective of RQ1 is to comprehensively evaluate the effectiveness of AutoLogger compared to state-of-the-art automated logging approaches. Automated logging involves multiple interconnected decisions that collectively determine the quality of generated logging statements. To systematically assess our approach's capabilities across these dimensions, we decompose this evaluation into three sub-questions:

RQ1.1: How effectively does AutoLogger address the whether-to-log decision?

RQ1.2: How well does AutoLogger perform on where-to-log and what-to-log tasks when logging is required?

**Table 2: Performance comparison on whether-to-log decision. The red and blue markers represent the best and second-best results.**

| Model | BA | Precision | Recall | F1 |
|---|---|---|---|---|
| Claude-sonnet-4 | 74.82 | 54.49 | 61.30 | 57.69 |
| Deepseek-r1 | 81.99 | 66.30 | 72.36 | 69.20 |
| Deepseek-v3.1 | 80.70 | 50.46 | 79.09 | 61.61 |
| GPT5 | 84.50 | 63.54 | 73.32 | 68.08 |
| Autologger w/o fine-tuning | 60.56 | 23.18 | 86.30 | 36.54 |
| Autologger | 96.92 | 99.47 | 93.95 | 96.63 |

RQ1.3: What is AutoLogger's end-to-end performance in realistic deployment scenarios?

*4.2.2 Approach.* For RQ1.1, we evaluate the whether-to-log decision using the complete test set, which contains both need-logging and no-need-logging methods. Since existing automated logging approaches do not explicitly address this fundamental decision, we compare AutoLogger against the LLM baselines described in Section 3.2.

For RQ1.2, we focus specifically on methods that require logging statements, using all positive instances from the test set. This evaluation encompasses both conventional approaches (i.e., UniLog, FastLog, and SCLogger) and LLM baseline based on Deepseek-v3.1. This sub-RQ us to assess AutoLogger's performance on the traditional automated logging tasks of location prediction and content generation when the logging requirement is already established.

For RQ1.3, we evaluate the complete end-to-end pipeline, where where-to-log and what-to-log predictions are only evaluated for methods correctly identified as requiring logging. This setup reflects realistic deployment scenarios where the system must first determine whether logging is needed before deciding on specific logging details. Based on the results from RQ1.2, we select the best-performing approach baseline (i.e., SCLogger) and the best-performing LLM baseline (i.e., Raw LLM with RAG) as comparison targets. Since SCLogger lacks whether-to-log capabilities, we augment it by combining with our backbone LLM (i.e., Deepseek-v3.1) to handle the whether-to-log decision.

*4.2.3 Result.* *RQ1.1.* Table 2 presents the performance comparison of different models on the whether-to-log decision task. AutoLogger achieves exceptional performance in determining whether-to-log, substantially outperforming all LLM baselines with a balanced accuracy of 96.92% and F1-score of 96.63%. Specifically, our fine-tuned model demonstrates a remarkable 27.43 percentage point improvement over the best-performing baseline Deepseek-r1, which achieves 69.20% F1. This superior performance stems from AutoLogger's near-perfect precision of 99.47%, effectively eliminating false positives while maintaining strong recall of 93.95%. The comparison with AutoLogger without fine-tuning reveals F1 of only 36.54% and precision of merely 23.18%, underscoring the critical importance of task-specific fine-tuning. The raw model suffers from severe overprediction despite achieving a high recall of 86.30%. Among the LLM baselines, GPT5 shows the most competitive performance with a balanced accuracy of 84.50% and F1 of 68.08%, yet it still falls significantly short of the capabilities of AutoLogger. These results demonstrate that fine-tuning is essential for accurate whether-to-log predictions, with AutoLogger achieving near-perfect precision while maintaining high recall.

*RQ1.2.* Table 3 compares the performance of different approaches on where-to-log and what-to-log tasks when evaluated exclusively on methods requiring logging. AutoLogger demonstrates consistent advantages across multiple dimensions, achieving the highest overall quality score with LLMJudge-Score of 1.3197, representing a 13.90% improvement over the best baseline SCLogger at 1.1587. For logging location prediction, AutoLogger achieves 61.30% PA, surpassing the best baseline Fastlog at 53.61% by 7.69 percentage points. In logging level prediction, AutoLogger attains 76.47% LA, slightly outperforming SCLogger at 75.84%. For variable selection, AutoLogger achieves the highest PMR of 59.22%, though SCLogger demonstrates superior Variable-F1 of 64.51% compared to 63.46%. In static text generation, AutoLogger excels in ROUGE metrics with ROUGE-L of 54.84%, though SCLogger achieves higher BLEU-4 scores of 35.71% versus our 30.35%. Overall, AutoLogger achieves superior or competitive performance across most logging dimensions when evaluated on methods requiring logging.

*RQ1.3.* Table 4 reports the end-to-end performance of different approaches in realistic deployment scenarios where whether-to-log predictions gate subsequent decisions. AutoLogger demonstrates substantial end-to-end advantages across all dimensions. Notably, AutoLogger achieves the highest LLMJudge-Score of 1.3646. This represents a 16.13% improvement over SCLogger at 1.1752. Building upon its exceptional whether-to-log performance with BA of 96.92% and F1 of 96.63%, AutoLogger achieves 57.21% PA, an 18.26% improvement over the best baseline configuration of SCLogger augmented with Deepseek-v3.1 at 48.37%. Beyond accurate where-to-log decisions, AutoLogger also demonstrates superior performance in what-to-log tasks, attaining 76.05% LA, 58.52% variable PMR, and 64.35% Variable-F1. While SCLogger exhibits marginally higher BLEU-4 scores at 35.04% versus 30.91%, AutoLogger surpasses it in ROUGE-L with 55.14% compared to 52.18%. These findings confirm that AutoLogger excels not only in determining logging necessity but also in generating high-quality logging statements across all dimensions in realistic end-to-end scenarios.

> **Answer for RQ1:** AutoLogger significantly outperforms state-of-the-art approaches across all automated logging dimensions, achieving 96.63% F1 in whether-to-log decisions and 16.13% improvement in logging statement quality.

## 4.3 RQ2: Contribution of Tools

*4.3.1 Motivation.* AutoLogger incorporates multiple specialized tools to enhance automated logging. Each tool is designed to address specific challenges in the logging task by providing contextual information or program analysis capabilities. However, the extent to which each tool contributes to the final performance remains unclear. Understanding these contributions is crucial for validating our design choices and identifying the most critical tools for practitioners seeking to adopt or extend our approach.

**Table 3: Performance comparison on where-to-log and what-to-log tasks for methods requiring logging. The red and blue markers represent the best and second-best results.**

| | Position | Levels | | Variables | | Texts | | | | Overall Quality |
|---|---|---|---|---|---|---|---|---|---|---|
| | PA | LA | AOD | PMR | F1 | BLEU-1 | BLEU-4 | ROUGE-1 | ROUGE-L | LLMJudge-Score |
| Proprietary Large Language Model | | | | | | | | | | |
| Raw LLM | 28.85 | 56.57 | 83.96 | 47.50 | 53.62 | 30.50 | 14.60 | 36.62 | 35.62 | 0.5697 |
| Raw LLM + RAG | 51.68 | 74.42 | 91.63 | 57.21 | 62.02 | 47.82 | 32.78 | 53.52 | 52.35 | 1.1514 |
| Existing Approach | | | | | | | | | | |
| Unilog | 51.44 | 72.34 | 91.07 | 52.55 | 58.47 | 44.36 | 27.02 | 50.79 | 49.61 | 1.1201 |
| Fastlog | 53.61 | 63.28 | 86.52 | 48.76 | 54.96 | 36.75 | 20.63 | 43.25 | 41.46 | 1.1034 |
| SCLogger | 52.88 | 75.84 | 92.11 | 55.03 | 64.51 | 49.36 | 35.71 | 54.13 | 53.37 | 1.1587 |
| Autologger | 61.30 | 76.47 | 91.86 | 59.22 | 63.46 | 49.56 | 30.35 | 55.95 | 54.84 | 1.3197 |

**Table 4: End-to-end performance comparison in realistic deployment scenarios. The red and blue markers represent the best and second-best results.**

| | Logging-or-not | | Position | Level | | Variables | | Texts | | Overall Quality |
|---|---|---|---|---|---|---|---|---|---|---|
| | BA | F1 | PA | LA | AOD | PMR | F1 | BLEU-4 | ROUGE-L | LLMJudge-Score |
| Raw LLM + RAG | 80.70 | 61.61 | 40.14 | 73.65 | 90.72 | 60.48 | 63.33 | 33.00 | 54.53 | 1.1672 |
| SCLogger | 80.70 | 61.61 | 48.37 | 74.97 | 91.58 | 52.54 | 61.97 | 35.04 | 52.18 | 1.1752 |
| Autologger | 96.92 | 96.63 | 57.21 | 76.05 | 91.70 | 58.52 | 64.35 | 30.91 | 55.14 | 1.3646 |

*4.3.2 Approach.* To assess the contribution of each component, we conduct an ablation study by removing individual tools from the complete AutoLogger system separately . Specifically, we evaluate four variants: (1) *AutoLogger without tool pool*, which removes the entire set of auxiliary tools available to agents; (2) *AutoLogger without similar case retrieval*, which eliminates the mechanism for retrieving relevant logging examples from the codebase; (3) *AutoLogger without backward slicing*, which removes the program analysis capability for tracing data dependencies; and (4) *AutoLogger without variable extractor*, which eliminates the specialized component for identifying and extracting relevant variables for logging. Each variant is evaluated on the same test set used in RQ1.2, focusing exclusively on methods requiring logging.

*4.3.3 Result.* Table 5 presents the ablation study results, revealing the contribution of each tool within the tool pool to AutoLogger's overall performance. The results demonstrate that all tools contribute positively to the system's effectiveness, with varying degrees of impact across different logging dimensions. Removing the entire tool pool causes the most severe performance degradation across all metrics, with PA dropping by 18.75 percentage points, LA decreasing by 14.89 percentage points, PMR declining by 22.50 percentage points, and ROUGE-L falling by 20.04 percentage points. This substantial decline underscores that the tool pool as a whole is indispensable for achieving high-quality automated logging.

Among individual tools, the similar case retrieval mechanism emerges as the most critical component. Its removal results in significant performance drops, with PA decreasing by 15.87 percentage points and ROUGE-L declining by 16.69 percentage points. This significant impact demonstrates that learning from existing logging patterns in the codebase is crucial for making informed logging decisions. The backward slicing analyzer contributes moderately but consistently across all dimensions. Its absence primarily affects logging level prediction with a 4.52 percentage point drop in LA and text generation with a 5.14 percentage point decline in ROUGE-L. These results indicate that understanding data flow and dependencies helps agents make more contextually appropriate logging decisions. The variable extractor shows the smallest but still meaningful impact, with its removal most notably affecting variable selection, leading to a 4.61 percentage point decrease in PMR. Notably, the sum of individual tool impacts does not equal the total impact of removing the entire tool pool, suggesting synergistic interactions among the tools where they complement each other's capabilities. These findings confirm that AutoLogger's superior performance stems from the effective integration of all tools, with similar case retrieval being particularly crucial for achieving high-quality automated logging.

> **Answer for RQ2:** All tools in AutoLogger design contribute positively to automated logging performance. The similar case retrieval is the most critical tool, showing performance drops of up to 16.69 point when removed.

## 4.4 RQ3: Generalizability

*4.4.1 Motivation.* The effectiveness of AutoLogger relies on two key model choices: the fine-tuned model for the Judger and the backbone LLM for MAS. A critical question is whether AutoLogger's strong performance is tied to our specific model selection or if the framework itself is robust and model-agnostic. We must investigate if the AutoLogger architecture provides value independent of the underlying models.

**Table 5: Ablation study results showing the contribution of each tool. The red subscripts indicate the performance change compared to the complete AutoLogger framework.**

| | Position | Level | | Variables | | Texts | |
|---|---|---|---|---|---|---|---|
| | PA | LA | AOD | PMR | F1 | BLEU-4 | ROUGE-L |
| Autologger | 61.30 | 76.47 | 91.86 | 59.22 | 63.46 | 30.35 | 54.84 |
| - *w/o tool pool* | 42.55$_{-18.75}$ | 61.58$_{-14.89}$ | 86.15$_{-5.71}$ | 36.72$_{-22.50}$ | 52.95$_{-10.51}$ | 15.08$_{-15.27}$ | 34.80$_{-20.04}$ |
| - *w/o similar case retrieval* | 45.43$_{-15.87}$ | 63.49$_{-12.98}$ | 87.17$_{-4.69}$ | 49.23$_{-9.99}$ | 55.65$_{-7.81}$ | 18.94$_{-11.41}$ | 38.15$_{-16.69}$ |
| - *w/o backward slicing* | 59.13$_{-2.17}$ | 71.95$_{-4.52}$ | 90.08$_{-1.78}$ | 55.28$_{-3.94}$ | 59.14$_{-4.32}$ | 26.52$_{-3.83}$ | 49.70$_{-5.14}$ |
| - *w/o variable extractor* | 60.57$_{-0.73}$ | 75.72$_{-0.75}$ | 90.33$_{-1.53}$ | 54.61$_{-4.61}$ | 61.30$_{-2.16}$ | 28.27$_{-2.08}$ | 52.07$_{-2.77}$ |

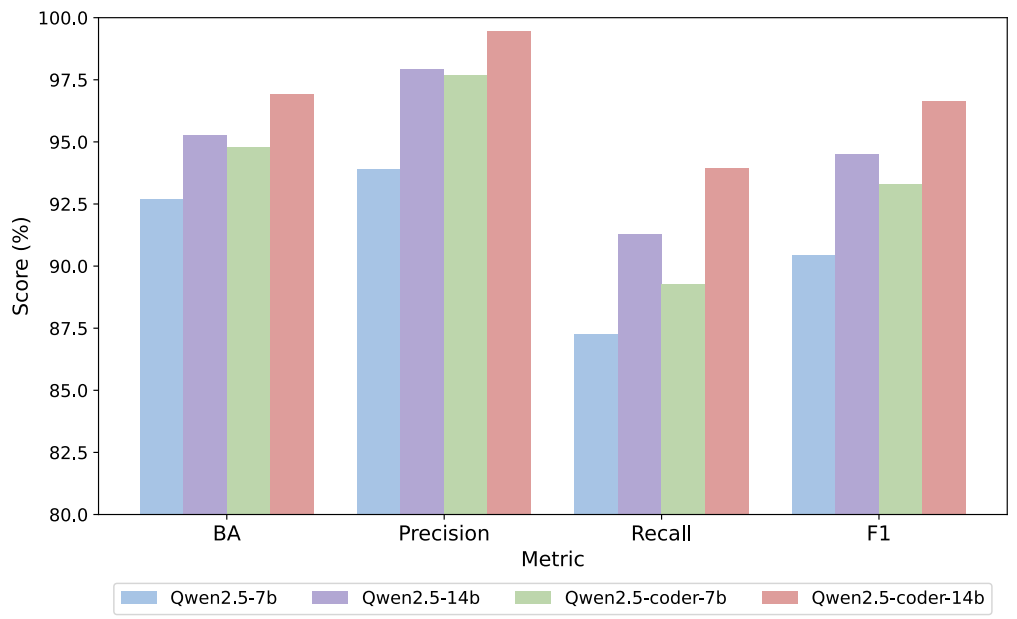

**Figure 2: Performance of different fine-tuned models on the whether-to-log task.**

*4.4.2 Approach.* To address RQ3, we conduct two separate experiments corresponding to the two stages of AutoLogger. First, we evaluate the generalizability of the Stage I Judger. We select four different models: Qwen2.5-7B, Qwen2.5-14B, Qwen2.5-Coder-7B, and Qwen2.5-Coder-14B. We fine-tune each of these models using the exact same dataset and LoRA configuration described in Section 2.2. We then evaluate each fine-tuned model on the whether-to-log task using the full test set. We compare their performance using BA, Precision, Recall, and F1-score.

Second, we assess the generalizability of the Stage II MAS framework. We select three different state-of-the-art backbone LLMs (i.e., GPT5, Claude-sonnet-4, and the default Deepseek-v3.1). For each LLM, we create two distinct experimental setups. In the first setup, we evaluate the raw LLM as a baseline, providing it with a single, comprehensive prompt to solve the logging task. In the second setup, we integrate the same LLM as the backbone for both the Locator and Generator agents within the full MAS framework. We then compare the performance of the raw LLM against our MAS framework on all where-to-log and what-to-log metrics. This comparison directly measures the performance uplift provided by our agent-based architecture.

*4.4.3 Result.* Figure 2 shows our fine-tuning strategy is highly generalizable, as all four models achieve strong F1-scores above 90.4%. The results reveal two clear trends: larger 14b models consistently outperform their 7b counterparts, and code-specialized models achieve better results than general-purpose models of the same size. This culminates in Qwen2.5-coder-14b delivering the best performance across all metrics, with 96.92% BA and 96.63% F1-score, confirming our Stage I approach is robust and validating our model choice for the Judger.

Table 6 demonstrates the generalizability of our MAS framework, which consistently improves the performance of all three tested backbone LLMs. The AutoLogger architecture provides the most significant boost in PA, with gains ranging from 18.65% to 22.75%. This enhanced location finding directly contributes to a substantial increase in overall quality, as measured by the LLMJudge-Score, which improved by 10.11% to 14.62% across all models. These results confirm that AutoLogger's specialized agent-based design adds significant value, regardless of the underlying LLM.

> **Answer for RQ3:** AutoLogger exhibits strong generalizability. Its fine-tuning strategy is effective across various models (all higher than 90.4% F1), and its MAS framework consistently improves the performance of different backbone LLMs, boosting overall quality scores by 10.11% to 14.62%.

# 5 Discussion

## 5.1 Usage of Tools by Agents

To understand the operational dynamics of our MAS, we analyzed the agent's tool-use patterns. We found that a complete AutoLogger execution requires only 2.39 tool calls on average. This low number demonstrates high efficiency, indicating the agents are decisive and do not engage in long exploratory chains. Furthermore, we measured the reliability of these calls and found that only 1.24% of tool invocations failed due to error arguments format from the LLM. While this low failure rate shows the model's strong adherence to tool schemas, its non-zero value is significant. It confirms that such failures are inevitable, validating the necessity of the robust fault-tolerance mechanisms, such as retries, which we designed into the MAS Orchestrator to ensure overall system reliability.

## 5.2 Limitations of AutoLogger

*5.2.1 Unsupport for Multiple Logging Statement Generation.* The current workflow of AutoLogger is designed to insert at most one logging statement per target method. While this design simplifies the generation task, it does not address scenarios where complex methods require multiple logging statements at different execution points. Future work could extend the MAS framework to overcome this limitation, for instance, by enabling the Locator to identify all suitable insertion candidates and empowering the Orchestrator to iteratively invoke the Generator for each identified location.

**Table 6: Performance comparison of base LLMs versus the AutoLogger framework.**

| Model | Approach | Position | Level | | Variables | | Texts | | Overall Quality |
|---|---|---|---|---|---|---|---|---|---|
| | | PA | LA | AOD | PMR | F1 | BLEU-4 | ROUGE-L | LLMJudge-Score |
| Claude-sonnet-4 | base | 56.01 | 78.54 | 92.38 | 62.23 | 64.72 | 34.29 | 57.50 | 1.3077 |
| | AutoLogger | 68.75 | 83.56 | 94.74 | 63.14 | 65.67 | 35.96 | 59.17 | 1.4426 |
| | Δ | ↑22.75% | ↑6.39% | ↑2.55% | ↑1.46% | ↑1.47% | ↑4.87% | ↑2.90% | ↑10.32% |
| Deepseek-v3.1 | base | 50.72 | 71.56 | 90.40 | 57.21 | 62.02 | 32.78 | 52.35 | 1.1514 |
| | AutoLogger | 61.30 | 76.47 | 91.86 | 59.22 | 63.46 | 30.35 | 54.84 | 1.3197 |
| | Δ | ↑20.86% | ↑6.86% | ↑1.62% | ↑3.51% | ↑2.32% | ↓7.41% | ↑4.76% | ↑14.62% |
| GPT5 | base | 52.88 | 79.55 | 93.30 | 57.73 | 64.25 | 38.11 | 59.06 | 1.2307 |
| | AutoLogger | 62.74 | 81.84 | 94.89 | 59.97 | 66.03 | 38.65 | 60.55 | 1.3551 |
| | Δ | ↑18.65% | ↑2.88% | ↑1.70% | ↑3.88% | ↑2.77% | ↑1.42% | ↑2.52% | ↑10.11% |

*5.2.2 Error Propagation in the Two-Stage Pipeline.* AutoLogger's two-stage architecture introduces a risk of error propagation. The hard-gating mechanism of the Judger implies that its classification errors are irreversible. A false positive will unnecessarily activate the resource-intensive MAS, potentially compelling the agents to generate low-quality logging statements and contributing to over-logging. A potential direction for future work is to implement a Reviewer agent, endowing the MAS agents with a veto capability to reject low-confidence or low-value generated logging statements.

## 5.3 Threat to Validity

*5.3.1 Internal Validity.* A significant threat to internal validity is the potential for data leakage. Our subject projects are prominent and their source code may have been included in the pre-training corpora of the LLMs we evaluated. This creates a risk of pre-training contamination, where a model's prior exposure to the test data could lead to an overestimation of its true performance. While we employed mitigation techniques, such as code reformatting and method encapsulation, these measures do not completely guarantee the elimination of this risk.

*5.3.2 External Validity.* Our empirical evaluation was conducted exclusively on three large-scale, mature Java projects within the distributed systems domain. These projects possess specific architectural patterns and established logging conventions which may not be representative of software developed in other languages. Furthermore, our evaluation follows a within-project setting where the training and test sets are derived from the same group of projects. While this design aligns with our goal of capturing project-specific logging conventions, it may overestimate performance when applying AutoLogger to a completely new project without prior fine-tuning (cross-project prediction). Future work will investigate the transferability of AutoLogger to unseen repositories.

# 6 Related Work

## 6.1 Automated Logging Statement Generation

Traditionally, the field of automated logging aims to determine both where-to-log and what-to-log. Early research treated these as separate problems, employing traditional machine learning with code features to predict logging locations [16, 45, 48] and information retrieval techniques to generate message content [5, 17, 19, 20]. The paradigm shifted with LANCE [23] and LEONID [22], where sequence-to-sequence models first offered a unified, end-to-end approach, treating log generation as a direct translation from a code snippet which lacks logging statement to a complete code snippet, thus combining the "where" and "what" decisions. Additionally, Xie et al. [39] introduced FastLog, an efficiency-focused tool capable of swiftly generating and inserting entire logging statements to streamline the end-to-end logging workflow.

More recently, the advent of LLMs has pushed the state-of-the-art forward substantially, owing to their vast pre-training on code and natural language. An extensive empirical study by Li et al. [14] systematically evaluated various LLMs on this task, establishing strong performance baselines and confirming their advanced code comprehension capabilities. Current research now focuses on augmenting these powerful models with deeper program context [39, 40]. For instance, Xu et al. [40] presented UniLog, grounded in the in-context learning framework of LLMs, which guides the model to generate appropriate logs by providing carefully selected examples within the prompt. Meanwhile, SCLogger [15] highlights the performance gains from enriching the model's input with inter-procedural information derived from static analysis, allowing the model to generate more contextually aware logs. As the variety of techniques has grown, the development of comprehensive evaluation benchmarks like AL-Bench [33] has also become essential, providing standardized benchmarks and dynamic evaluation methodologies to rigorously compare the performance of different models.

Different from this prior work, AutoLogger introduces two fundamental innovations. First, it is the first to systematically address the neglected *whether-to-log* decision, which existing tools universally ignore. Second, instead of using a monolithic model, AutoLogger employs a MAS that decomposes the generation task. This allows specialized agents to use program analysis tools, grounding their reasoning in verifiable facts to mitigate hallucination.

## 6.2 LLM-Based Multi-Agent Application in SE

Recent research in SE leverages LLM-based MASs to automate development by simulating human team dynamics [3, 21]. The core principle of this paradigm is the decomposition of complex workflows into discrete sub-tasks, which are then assigned to agents assuming specialized roles. Collaboration among these agents is governed by structured communication protocols. For instance, ChatDev [29] constructs a virtual company where agents handle the entire software development lifecycle. For more targeted applications, LLM4FL [30] employs a trio of agents to perform fault localization, while MAGIS [34] orchestrates a team of agents to autonomously resolve software maintenance issues from platforms like GitHub.

Our work applies this MAS paradigm in a novel way. While existing systems like ChatDev target broad development workflows, AutoLogger is the first to design a specialized MAS for the *automated logging* task. Its hybrid architecture, which combines a fine-tuned classifier with the agent team, and its dedicated pool of program analysis tools are tailored specifically to solve the unique data-flow and stylistic challenges of logging.

# 7 Conclusion

Existing automated logging tools often neglect the crucial whether-to-log decision and struggle with the composite nature of logging using monolithic models. To address this, we proposed AutoLogger, a novel hybrid framework combining a fine-tuned Judger classifier for logging necessity with a specialized MAS for grounded generation. Our evaluation shows AutoLogger achieves a 96.63% F1-score on the whether-to-log task and improves end-to-end logging quality by 16.13% over the strongest baseline. This work demonstrates the significant benefits of task decomposition and component specialization for building effective automated logging solutions.

# Acknowledgments

The work described in this paper was supported by two grants from the Research Grants Council of the Hong Kong Special Administrative Region, China: (1) No. CUHK 14209124 of the General Research Fund, and (2) No. SRFS2425-4S03 of the Senior Research Fellow Scheme.